\documentclass[sigconf,
                    ]{acmart}  

\AtBeginDocument{%
  }

\copyrightyear{2024}
\acmYear{2024}
\setcopyright{rightsretained}
\acmConference[L@S '24]{Proceedings of the Eleventh ACM Conference on Learning @ Scale}{July      18--20, 2024}{Atlanta, GA, USA}
\acmBooktitle{Proceedings of the Eleventh ACM Conference on Learning @ Scale (L@S '24), July      18--20, 2024, Atlanta, GA, USA}
\acmDOI{10.1145/3657604.3662028}
\acmISBN{979-8-4007-0633-2/24/07}

\usepackage{hyperref}
\usepackage[framemethod=TikZ]{mdframed}
\usepackage{lipsum}
\usepackage{xcolor}
\definecolor{myblue}{HTML}{4E7AC7}
\usepackage{float}
\usepackage{multirow}
\usepackage{enumerate}

\begin{document}

\title[Differential Course Functioning]{Gaining Insights into Group-Level Course Difficulty  via~Differential~Course~Functioning}

\author{Frederik Baucks}
\authornote{Both authors contributed equally to this research.}
\affiliation{
  \institution{Ruhr-Universität Bochum}
  \streetaddress{Universitätsstraße 150}
  \city{Bochum}
  \country{Germany}
  }
\email{frederik.baucks@ini.rub.de}

\author{Robin Schmucker}
\authornotemark[1]
\affiliation{
    \institution{Carnegie Mellon University}
    \streetaddress{5000 Forbes Ave}
    \city{Pittsburgh}
    \state{PA}
    \country{USA}
}
\email{rschmuck@cs.cmu.edu}

\author{Conrad Borchers}
\affiliation{
    \institution{Carnegie Mellon University}
    \streetaddress{5000 Forbes Ave}
    \city{Pittsburgh}
    \state{PA}
    \country{USA}
}
\email{cborcher@cs.cmu.edu}

\author{Zachary A. Pardos}
\affiliation{
    \institution{University of California, Berkeley}
    \city{Berkeley}
    \state{CA}
    \country{USA}
}
\email{pardos@berkeley.edu}

\author{Laurenz Wiskott}
\affiliation{
  \institution{Ruhr-Universität Bochum}
  \streetaddress{Universitätsstraße 150}
  \city{Bochum}
  \country{Germany}
  }
\email{laurenz.wiskott@ini.rub.de}

\renewcommand{\shortauthors}{Baucks et al.}

\begin{abstract}
Curriculum Analytics (CA) studies curriculum structure and student data to ensure the quality of educational programs. One desirable property of courses within curricula is that they are not unexpectedly more difficult for students of different backgrounds. While prior work points to likely variations in course difficulty across student groups, robust methodologies for capturing such variations are scarce, and existing approaches do not adequately decouple course-specific difficulty from students' general performance levels. The present study introduces Differential Course Functioning (DCF) as an Item Response Theory (IRT)-based CA methodology. DCF controls for student performance levels and examines whether significant differences exist in how distinct student groups succeed in a given course. Leveraging data from over $20{,}000$ students at a large public university, we demonstrate DCF's ability to detect inequities in undergraduate course difficulty across student groups described by grade achievement. We compare major pairs with high co-enrollment and transfer students to their non-transfer peers. For the former, our findings suggest a link between DCF effect sizes and the alignment of course content to student home department motivating interventions targeted towards improving course preparedness.
For the latter, results suggest minor variations in course-specific difficulty between transfer and non-transfer students. While this is desirable, it also suggests that interventions targeted toward mitigating grade achievement gaps in transfer students should encompass comprehensive support beyond enhancing preparedness for individual courses.
By providing more nuanced and equitable assessments of academic performance and difficulties experienced by diverse student populations, DCF could support policymakers, course articulation officers, and student advisors.
\end{abstract}




\maketitle
\section{Introduction}
\label{sec:introduction}

Curriculum Analytics (CA) employs data-driven methodologies to identify and understand factors that affect students' experience and contribute to student outcomes such as grades (e.g., \cite{baucks21mitigating, baucks24lak}), dropout (e.g.,   \cite{salazar2021curricular,aina2022determinants}), and time to degree (e.g., \cite{molontay2020characterizing, baucks2022simulating}).
CA insights support stakeholders and policymakers, informing targeted interventions benefiting learning outcomes, equity, and fairness~\cite{Uttamchandani2022:Introduction,Gardner2023:Cross}. Previous CA research employed pass rates and course grades as measures to identify systematic variations in course difficulty across student groups (e.g.,~\cite{Johnson2005:Academic,Bronkema2017:Residential}). While these studies highlight undesirable course difficulty differences between student groups, the employed measures of difficulty need to be interpreted with care as they do not adequately decouple \textit{course-specific difficulty} from students' \textit{general performance levels}. This can lead to confounding in estimates of differences between groups~\cite{baucks24lak}. Achieving such decoupling is crucial to evaluate if and understand how \textit{individual} courses pose disparate challenges for different students groups (e.g., transfer/non-transfer). Adequate approaches to modeling group-level difficulty differences could inform interventions mitigating related inequities.

This paper introduces Differential Course Functioning (DCF) as an Item Response Theory (IRT)-based CA methodology that controls for students' individual performance levels. DCF assesses how distinct groups of students succeed in individual courses. IRT was first developed for high-stakes assessment and employs statistical techniques to measure latent traits of test takers and difficulty of test items \cite{Ayala2013:Theory}. IRT models responses to multiple test items (e.g., multiple-choice questions) by assuming that a student's response to a particular item can be explained via a probabilistic link between the student's trait and the item's difficulty. In CA research, recent studies have successfully adopted IRT-based methodologies for the analysis of student course grades (e.g., \cite{Bacci2017:Evaluation,baucks24lak}), where courses are analogous to test items. By decoupling student traits and course difficulty, IRT can yield valuable CA insights. For example, by monitoring course difficulty variations over time, IRT revealed systematic shifts in difficulty levels during the COVID-19 pandemic~\cite{baucks24lak}. One major limitation of IRT is its assumption that course difficulty is \textit{constant} for all students regardless of their background. Existing evidence (e.g.,~\cite{zlatkin2019ethics,fematt2021identifying,ishitani2008transfers}) challenges this assumption. 
%

Some groups of students might experience disparate outcomes in courses even after controlling for the course's difficulty and the student's general performance level. 
In this context, transfer status and student major are two particularly relevant group attributes. For example, the University of California system aims to enroll one transfer student for every two freshmen \cite{UC2021Accountability}. 
However, the transfer process can cause students to loose course credits~\cite{taylor2017multiple} and to face additional challenges related to the change in learning environment--a "transfer shock" phenomena \cite{ishitani2008transfers,taylor2017multiple}. On the other side, assessing potential differences between students pursuing different majors is challenging because individual majors often share courses. Prior knowledge related to the major and teaching styles of respective home departments may influence how students experience these shared courses \cite{hailikari2008relevance}. 

In the context of high-stakes assessments, differential item functioning (DIF) is a technique used to assess and mitigate inequities in test difficulty associated with to group-level attributes~\cite{Van2013:Handbook}. Relatedly, this work explores differential \textit{course} functioning (DCF) as a CA methodology and demonstrates its ability to detect disparate difficulties in undergraduate courses across student groups described by grade achievement, by analyzing data from over 20,000 students at a large public university. Our contributions include:

\begin{description}
\item [DCF as CA methodology:] We extend IRT-based CA methodologies to assess relationships between grade achievement and group attributes. DCF enables us to critically examine whether significant differences exist in how distinct student groups succeed in a specific course, particularly after accounting for their overall performance level.
\item [Home Department DCF:] By analyzing course-specific difficulty factors relating to departmental affiliations, we uncover evidence suggesting that courses closely aligned with the content of a student's home department tend to be easier for those majors. The magnitude of these DCF effects offers valuable insights for student advising and developing targeted interventions to improve course preparedness.
\item [Transfer Status DCF:] Our analysis revealed no substantial differences in the treatment of transfer and non-transfer students related to course-specific difficulty factors. This finding suggests that efforts to bridge the gap in A-grade achievement should encompass comprehensive support beyond enhancing preparedness for individual courses.
\end{description}

\section{Related Work}
\label{sec:related_work}

\subsection{Curriculum Analytics}
\label{subsec:curriculum_analytics}

Three main methodologies for data-driven analysis have emerged in the field of CA: (i) the use of process mining techniques (e.g., \cite{brown2018taken,wagner2022combined,martinez2023evaluation}), (ii) the simulation of educational processes (e.g., \cite{molontay2020characterizing,mceneaney2022curriculum,baucks2022simulating}), and (iii) the adoption of predictive methods based on curriculum structure (e.g., \cite{slim2014employing,backenkohler2018data,pardos2020university}). Firstly, process mining is used to construct models of the educational process, focusing on the sequence of interactions with different curriculum elements (e.g., courses). Secondly, building on process-mining, simulation techniques are used to assess the impact of potential changes within the curriculum. Finally, predictive models have been developed that focus on estimating students' future academic performance and providing personalized recommendations for students' curricular pathways.

In recent years, questions related to equity and fairness in education emerged as central themes in the community~\cite{fischer2020mining,Yu2021:Should,Uttamchandani2022:Introduction}. Related CA research focuses on identifying and understanding factors connected to disparities in educational outcomes across student groups (e.g., procrastination~\cite{sabnis2022large} and housing~\cite{Bronkema2017:Residential}). Outcome measures commonly considered in this context are course pass rates \cite{wagner2022combined, molontay2020characterizing,baucks2022simulating} and student GPA \cite{Mendez2014:Curricular, srivastava2024curriculum}. However, both of these measures do not adequately decouple the relationships between outcomes, variations in student cohort performance levels, and variations in course difficulty~\cite{Hansen2019:Estimating,baucks24lak}. The DCF methodology introduced in this study accounts for these factors and provides a more nuanced measure explicitly designed to assess disparate difficulties in individual courses associated with group-level attributes.

\subsection{IRT in Curriculum Analytics}
\label{subsec:irt_in_ca}

Recent CA research explored the application of IRT-based methodologies to evaluate course difficulty in university degree programs~\cite{Bacci2017:Evaluation,haas2023bayesian,baucks24lak}. The original motivation for IRT was to overcome limitations of classical test theory in high-stakes assessments. Most importantly the lack of comparability of scores from different tests and the dependency of item parameters on the test taker cohort~\cite{Ayala2013:Theory}. In CA, IRT treats courses analogous to test items.

Related research includes the multidimensional latent class IRT model from Bacci et al.~\cite{Bacci2017:Evaluation} which assigns first-year students in Italy to different performance groups based on exam enrollment and grade data. As part of their work, Bacci et al.~\cite{Bacci2017:Evaluation} pointed out variations in course difficulty between student groups related to lecturer assignments which were based on students' last names. Baucks et al.~\cite{baucks24lak} adapted IRT to assess temporal variations in course difficulty at a university in Germany over several years. Studying data from two STEM majors, they revealed fluctuation in course difficulty. This paper extends the methodology from Baucks et al.~\cite{baucks24lak} and focuses on examining how group-level attributes (i.e., home department and transfer status) relate to variations in course difficulty across student groups by leveraging data from $6$ undergraduate programs at a large public university in the United States.

\subsection{Differential Item Functioning}
\label{subsec:irt_dif}

One limitation of the IRT framework is its assumption that the difficulty of individual courses (items) is \textit{equal} for all students regardless of their personal background. However, in the real world, experienced difficulties can vary between groups of students (e.g., due to language proficiency~\cite{fematt2021identifying}). To mitigate inequities in student assessments, \textit{differential item functioning} (DIF) has been introduced as a methodology that provides an explicit measure of difficulty variations between different groups~\cite{Ayala2013:Theory}.

In standardized testing, DIF analysis is used to detect and mitigate disparate difficulties in assessment items. One prominent example is the Programme for International Student Assessment (PISA) directed by the Organization for Economic Cooperation and Development (OECD). PISA assesses 15-year-old students' knowledge in a range of subjects and conducts tests in 79 countries worldwide (as of 2018)~\cite{schleicher2019pisa}. DIF analysis is used to control for difficulty variations related to student gender in subject-specific items and item formats (e.g., \cite{le2009investigating,shear2023gender, cheema2019cross,khorramdel2020examining}, and to assess differences between native and immigrant students, for example, in reading proficiency~\cite{da2012differential}.

In the context of curriculum analytics, the present study takes inspiration from DIF analysis and introduces \textit{differential course functioning} (DCF) as a statistical model to detect and quantify variations in course difficulty related to group-level attributes. 
We showcase DCF's ability to detect inequities in difficulty described by student grade achievement by comparing groups of: Firstly, major pairs with high co-enrollment as they may indicate departmental differences in teaching and learning \cite{hailikari2008relevance}. Secondly, transfer to non-transfer students as DCF detections can be indicative for courses in which divergent prior knowledge or learning conditions exist \cite{taylor2017multiple,ishitani2008transfers,taylor2017multiple}. 

\section{Methodology}
\label{sec:methodology}

\subsection{Item Response Theory}
\label{irt_methodology}

In CA, IRT uses course grade data to infer for each student the latent trait value, which best explains the likelihood of the student surpassing the specified grade threshold in each course (e.g., passing or median grade), hereafter called achievement grade (AG)~\cite{Ayala2013:Theory}. IRT can model traits via one or more dimensions. In multidimensional cases, dimensions can represent distinct skills (e.g., language and logic proficiency). To select the IRT model most suitable for our data, we follow prior work applying IRT to similar data~\cite{baucks24lak}: (i) Fit one- and multidimensional IRT models; (ii) choose the best-fitting model; (iii) assess whether the data meets IRT's assumptions.

The relationship between student trait values and course achievement rates (ARs) is modeled by fitting a sigmoid function for each course, known as the item response function (IRF) (i.e., the green curve in Figure~\ref{fig:dcf}). The inverse image (x-axis) of the IRF is the student’s trait value, and the image (y-axis) is the student’s probability of reaching the AG in a specific course.

Let $C$ and $S$ be the sets of courses and students, respectively. For course $c \in C$, the potentially multidimensional location of its IRF on the $x$-axes is defined by the course location vector $\boldsymbol{\delta}_c \in \mathbb{R}^n$. 
The slope of the IRF characterizes the course's discrimination property denoted as $\boldsymbol{\alpha}_c \in \mathbb{R}^n$. Given student trait $\boldsymbol{\theta}_{s \in S} \in \mathbb{R}^n$, course location $\boldsymbol{\delta}_c$, and course discrimination $\boldsymbol{\alpha}_c$, student $i$'s probability of reaching the AG in course $j$ is defined as
\begin{align}\label{IRF}
    \mathbb{P}(X_{s,c} =1 \,|\, \boldsymbol{\theta}_s,\, \boldsymbol{\alpha}_c,\, \boldsymbol{\delta}_c) &\,= \frac{1}{1+ e^{-\langle \boldsymbol{\alpha}_c, \boldsymbol{\theta}_s - \boldsymbol{\delta}_c \rangle}},
    %
\end{align}
where $X_{s,c} \in \{ 0,1 \}$ is the dichotomous response of student $s \in S$ to course $c \in C$ and $\langle\cdot,\cdot\rangle$ denotes the Euclidean inner product. The position on the $x$-axes where the IRF has its largest slope indicates the course difficulty. We define the single-dimensional course difficulty $\Delta_c$ of course $j$ as:
\begin{align}\label{Difficulty}
    \Delta_c & = \frac{\langle \boldsymbol{\alpha}_c, \boldsymbol{\delta}_c \rangle}{\| \boldsymbol{\alpha}_c \|_2} \in \mathbb{R}.
\end{align}
    
    The IRT model defined by Eq.~\ref{IRF} can be fitted via maximum likelihood estimation based on course grade data. When choosing model dimension $n = 1$ and setting $\alpha_c=1$ (for all $c \in C$), the resulting \textit{unidimensional} course locations $\delta_c \in \mathbb{R}$ equal the course difficulty estimates ($\delta_c = \Delta_c$). Focusing on optimizing the unidimensional course locations $\delta_c$ while keeping all $\alpha_c = 1$ is known as Rasch or 1-parameter logistic model (1PL) \cite{Ayala2013:Theory}. When the constraints on discrimination parameters $\alpha_c$ are removed, allowing them to vary freely, this formulation is known as Birnbaum or 2-parameter logistic model (2PL) \cite{Ayala2013:Theory}. 
    The Rasch and Birnbaum model assume that the grade data is best explained by single-dimensional traits, which is also known as unidimensionality assumption \cite{Ayala2013:Theory}. To assess whether this assumption is met, we compare fit to multidimensional models.
    As an extension of the Birnbaum model, the multidimensional IRT model (MIRT)~\cite{Chalmers2012:Mirt} uses multidimensional parameter vectors ($n>1$) to describe course difficulty, discrimination, and student traits. 
    The $n$-dimensional variant of the IRT model is referred to as 2PL-nDIM, e.g. $2$-dimensional $\sim$ 2PL-2DIM.

    Importantly, IRT is designed to analyze and explain student performance data in \textit{retrospective}, focusing on the fit of course and student parameters to past data. Although these parameters could be used for forecasting purposes, such as predicting the difficulty of future courses, this study focuses solely on examining past performance data to generate insights on the educational process. Consequently, our analysis conducts model selection via an information criterion that balances between model accuracy and model complexity via an \textit{in-sample} evaluation (details Section~\ref{model_assumptions}). 

    \subsection{IRT Assumptions and Measure Properties}
    
    \subsubsection{Verifying Model Assumptions} 
    \label{model_assumptions}
    IRT makes two assumptions about the data: First, the dimensionality assumption requires that the latent traits for the courses and students of a given dimension are, in fact, a good representation of the data; second, local independence postulates that, given a student's underlying IRT trait level, responses to individual courses are independent of each other. To assess \textit{dimensionality}, we consider IRT models with one to three dimensions~(see Section \ref{irt_methodology}) with nested parameter spaces. This enables us to employ the Bayesian Information Criterion (BIC) to determine the best trade-off between model complexity (parameter number) and model fit (log-likelihood) to decide how many dimensions are appropriate \cite{kang2007irt}. For \textit{local independence}, we examine correlations between regression residuals (the differences between observed and expected responses) using Yen's Q3~\cite{Yen1993:Scaling} to estimate pairwise dependencies between courses. If the difference in Q3 score between a pair exceeds a threshold of 0.2 from the average Q3 score of all pairs, local independence may be compromised~\cite{christensen2017critical}.
\subsubsection{Validity and Reliability of Measures} 

We evaluate validity and reliability to confirm accuracy and consistency of the estimated student and course parameters, ensuring they effectively represent the constructs of interest--student performance and course difficulty. 

For \textit{validity}, we evaluate \textit{concurrent validity} (i.e., the agreement between measures), by comparing IRT trait values with variables expected to capture similar attributes. We assess correlations between the student trait captured by the absolute value of the IRT trait vector $\|\theta_s \|_2$ and student GPA as well as correlations between the IRT course difficulty $\Delta_c$ and the course AR. Similar to GPA adjustment research (e.g.,~\cite{Hansen2019:Estimating, young1990adjusting}), we expect high correlations to indicate that we capture the targeted constructs.

For \textit{reliability}, we assess the consistency of fitted trait values via split-half testing, yielding a measure of \textit{internal consistency} reliability~\cite{Ayala2013:Theory}. Split-half testing partitions the data into two disjoint subsets to fit two separate models. Afterwards, it employs the Pearson correlation coefficient to measure consistency in trait values between the two different models. Following prior work~\cite{baucks24lak}, we partition the data in two different ways: (i) In the random partitioning, we split each student's course grades at random into two disjoint subsets; (ii) In the time-dependent partitioning, we divide students' course grades into earlier and later halves of their studies, thereby addressing IRT's assumption of time-invariant trait values.

\subsection{Differential Course Functioning}

\begin{figure}
    \centering
    \includegraphics[width = 0.95\columnwidth]{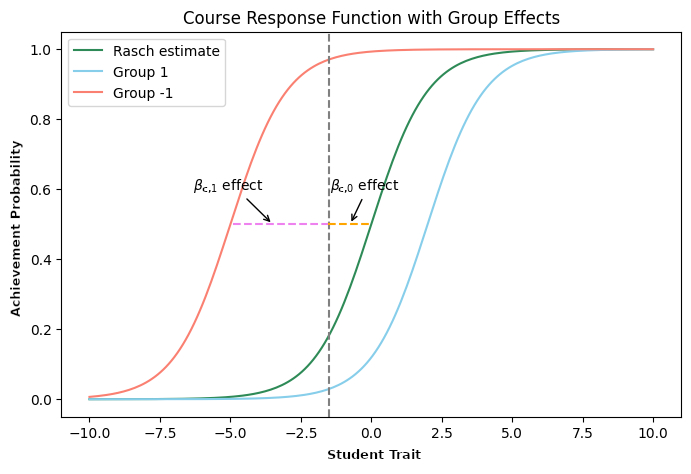}
    \caption{DCF model. The green sigmoid function indicates the response curve of a Rasch model fitted for all students. The red and blue curves indicate group-specific course responses (red $\sim -1$, blue $\sim 1$) that exhibit asymmetric offsets from the Rasch model. Parameter $\beta_{c,0}$ is the intercept of the logistic regression model and captures the distance between the Rasch IRF and the position on the x-axis from which both group IRFs are equidistant. This is the reference point for the group-specific DCF measure $\beta_{c,1}$ (difficulty disparity).} 
    \label{fig:dcf}
    \Description[]{long description}
\end{figure}

DIF methodologies quantify disparate difficulties related to group-level attributes. Statistical tests assess whether the data meets IRT's assumption that student performance is solely dependent on the trait measures. For example, a business administration student that takes an advanced statistics course might face additional challenges compared to a home department student due to limited course preparedness. For CA, we evaluate differential course functioning (DCF) as follows: For each course, we perform a second regression assessing potential differential functioning between two groups of students (e.g., transfer vs. non-transfer): 
\begin{align}\label{dcf_lr}
    \text{logit}(\mathbb{P}(X_{s,c} = 1 | \, \theta_s, \delta_c)) =  \beta_{c,0} + \beta_{c,1} g_s + (\theta_s - \delta_c).
\end{align}
Here, the logit function is the inverse of the sigmoid $1/(1+$~$e^{-x})$, $\theta_s$ is the trait of student $s \in S$ and $\delta_c$ is the difficulty of course $c \in C$ both fitted by the \textit{initial} Rasch model, $g_s \in \{-1, 1\}$ is the DCF group encoding $\beta_{c,1}$ is the DCF effect, and $\beta_{c,0}$ is the DCF intercept. Therefore, $\beta_{c,0}$ and $\beta_{c,1}$ are the \textit{only} parameters fitted in this second regression. Finding a DCF effect indicates that the course exhibits systematic variations in difficulty between the groups. A negative $\beta_{c,1}$ parameter suggests that students in the $g_s = -1$ group find course $c$ \textit{easier} than students in the $g_s = 1$ group (Figure~\ref{fig:dcf}). 

If we compare the red and blue sigmoid functions in Figure\ref{fig:dcf}, we observe that a DCF effect $\beta_{c,1}$ leads to different variations in AG probabilities between the considered groups depending on the individual student's trait value $\theta_s$. This is due to the non-linearity of the sigmoid function. The DCF effect value $\beta_{c,1}$ in Eq.~\ref{dcf_lr} is captured as a logit and affects the offset of the sigmoid function independent of course-specific difficulty and student trait values. This makes it non-trivial to compare a DCF effect in one course to a DCF effect in another. 
To make these difference more interpretable, our examination of DCF effects converts DCF logits into differences in achievement grade probabilities. 
For this, we take the average student trait value of the students independent of group assignment and calculate their achievement grade probabilities to make the DCF effect comparable between courses.

This study models DCF for Rasch models only. Rasch models do not involve a slope parameter or multidimensional trait or difficulty values (Section \ref{irt_methodology}). Consequently, we always assess whether the course grade data meet the Rasch model assumptions—which was the case for all but one of the considered majors (Table \ref{tab:selection_results}).

    \subsection{DCF Effect Interpretation} \label{dcf_vs_pr}
        For easier differentiation, we denote the group with DCF encoding $g_s = -1$ as $\boldsymbol{G}_1$ and the group with encoding $g_s = 1$ as $\boldsymbol{G}_2$.
        
        Achievement rates (ARs), e.g., pass rates, are often used to assess the difficulty of courses in educational curricula \cite{Mendez2014:Curricular}. A recent study highlighted how course pass rates must be interpreted with caution because pass rates are affected by \textit{student performance} level and \textit{course difficulty}~\cite{baucks24lak}. That is, if a difference in ARs is detected between two courses, the difference could be due to either factor or a combination of both. When identifying courses that pose disparate difficulties for two distinct groups, student performance and course difficulty need to be accounted for~\cite{baucks21mitigating}. The AR counterpart of DCF computes the AR difference ($\text{AR}_\Delta$), as the gap in ARs between the two groups ($\text{AR}_\Delta = \text{AR}_{G_1}-\text{AR}_{G_2}$), for example as the difference in A-grade achievement rates. The AR difference assumes uniformity across all students in a group. In contrast, DCF can capture student trait-dependent effects, i.e., shifts in the sigmoid as in Figure~\ref{fig:dcf}, and can assesses more nuanced non-uniform effects on the probability of reaching the AG across varying student trait levels.
        
\begin{figure}\flushleft
  
  \tikzset{
    level 1/.style = {level distance=0.cm, sibling distance=3.5cm},
    level 2/.style = {level distance=3.5cm, sibling distance=2cm},
    bag/.style = {text width=4cm, text centered,  inner sep=1pt},
    end/.style = {circle, minimum width=3pt,fill, inner sep=0pt}
  }
  \begin{tikzpicture}[grow=right, sloped, scale=0.625]
    \tikzset{frontier/.style={distance from root=150pt}}

    \node {}
    child {
      node[bag] (C) {$\text{AR}_{G_1} \neq \text{AR}_{G_2}$}        
      child {
        node[end, label=right:
        {\shortstack{Pot. varying trend:\\(IV) DCF $\in \mathbb{R},\, \text{AR}_\Delta \neq 0$}}] {}
        edge from parent
        node[above] {}
        node[below]  {$\theta_{G_1} \neq \theta_{G_2}$}
      }
      child {
        node[end, label=right:
        {\shortstack{Same trend:\\(III) $\text{DCF} \neq 0,\, \text{AR}_\Delta \neq 0$}}] {}
        edge from parent
        node[above] {$\theta_{G_1} = \theta_{G_2}$}
        node[below]  {}
      }
      edge from parent 
      node[above] {}
      node[below]  {}
    }
    child {
      node[bag] (B) {$\text{AR}_{G_1} = \text{AR}_{G_2}$}        
      child {
        node[end, label=right:
        {(II) $\text{DCF} \neq 0,\, \text{AR}_\Delta = 0$}] {}
        edge from parent
        node[above] {}
        node[below]  {$\theta_{G_1} \neq \theta_{G_2}$}
      }
      child {
        node[end, label=right:
        {(I) $\text{DCF} = 0,\, \text{AR}_\Delta = 0$}] {}
        edge from parent
        node[above] {$\theta_{G_1} = \theta_{G_2}$}
        node[below]  {}
      }
      edge from parent 
      node[above] {}
      node[below]  {}
    };
  \end{tikzpicture}
      \caption{Possible relationships between AR differences ($\text{AR}_{G_1}$, $\text{AR}_{G_2}$) and DCF for groups ($G_1$, $G_2$) with identical and deviating student traits values ($\theta_{G_1}$, $\theta_{G_2}$). DCF offers a more nuanced measure of academic performance to gain insights into difficulties experienced by diverse student populations.}
      \label{fig:dcf_pr_cases}
      \Description[]{long description}
    \end{figure}
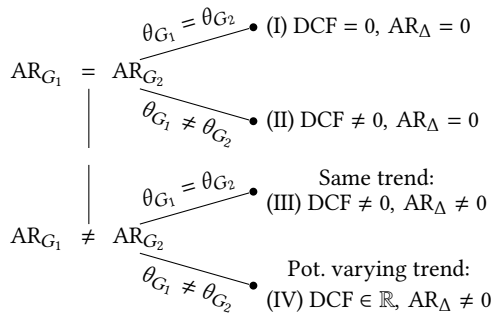

        Figure~\ref{fig:dcf_pr_cases} illustrates the possible cases for the effects of DCF and $\text{AR}_\Delta$ under different mean ARs and IRT student traits in groups $G_1$ and $G_2$ that can be identified by our methodology. In all cases (II-IV) except the null effect case (I)--where both effects are $0$--we can observe how DCF and $\text{AR}_\Delta$ effects function differently.
        The four cases illustrated in Figures~\ref{fig:dcf_pr_cases} and~\ref{fig:dcf_vs_pr_res} can be interpreted as follows:
        \begin{enumerate}[(I)]
            \item \textbf{Null Effect}: No evidence for disparate outcomes due to student trait differences ($\theta_\Delta$) or effects of DCF.
                \item \textbf{$\boldsymbol{\theta_\Delta}$}: Groups with \textit{varying} student trait levels achieve \textit{equal} outcomes due to disparate difficulties. 
                \item \textbf{DCF}: Groups with \textit{equal} student trait levels achieve \textit{disparate} outcomes due to disparate difficulties.
                \item \textbf{$\boldsymbol{\theta_\Delta}$+DCF}: Groups with \textit{varying} student trait levels achieve \textit{disparate} outcomes due to student trait differences and DCF.
        \end{enumerate}
    
    In the cases where latent student trait values of the two groups differ, course ARs will be confounded by the \textit{overall} performance differences between the participating students. By accounting for these factors, DCF promotes a more accurate and nuanced measure of academic performance, yielding insights into \textit{course-specific} difficulties experienced by students of different backgrounds. 

\subsection{Verifying Robustness of DCF Effects}
\label{simulation_and_bh}

To generate robust insights, we statistically control DCF effect ($\beta_{c,1}$) detection in two ways: First, using Benjamini-Hochberg adjustment~\cite{thissen2002benjamini} and likelihood ratio testing, we control the false discovery rate (FDR) (i.e., the expected ratio between false positives and the sum of false positives and true positives) under multiple tests (one for each course). Second, we conduct a simulation study to estimate the true positive rate (TPR) (i.e., the ratio between true positives and the sum of true positives and false negatives) of the DCF detection process by performing power analysis. 

\subsubsection{Accounting for Multiple Testing}

To evaluate whether the detected DCF effects $\beta_{c,1}$ are statistically significant, we assess regression model fit by comparing the likelihood of the Rasch model to our extended model for DCF detection (Eq. \ref{dcf_lr}) by using a likelihood ratio test~\cite{cohen1996investigation}.

When controlling for a statistical test's false positive rate (FPR) (i.e., the ratio between false positives and the sum of false positives and true negatives), we typically employ a \textit{p}-value threshold of $0.05$. However, when performing multiple statistical tests (i.e., one test for each course), the chance of falsely discovering an effect increases with each additional test. To account for this, we employ the Benjamini-Hochberg adjustment. This method works by ordering the individual \textit{p}-values from the different tests and then adjusting them based on their order and the total number of tests. We determine the appropriate Benjamini-Hochberg threshold to ensure a FDR of $0.05$.

\subsubsection{Statistical Power Simulation}
We evaluate the statistical power of our DCF detection method via simulation study~\cite{Jodoin2001:Evaluating}, that is, we obtain an estimate of the TPR under different DCF effect sizes ($\beta_{c,1} \in [0.1, 0.4]$) and consider varying population sizes ($|G_1| = |G_2| \in [50, 550]$) with two equally sized groups.

First, we sample Rasch student trait and $50$ course difficulty parameters from the normal distribution--similar to our smallest dataset with $46$ courses. Second, we simulate the student response data using the AR probabilities induced by the Rasch model. 
Third, we introduce DCF for one course by shifting the corresponding course response function defined by the Rasch model according to the students encoded in two groups. We assign each student to exactly one group and encode them with the labels $-1$ and $1$. 
Finally, we report the statistical power of the DCF method as the ratio of significant effects found among $1{,}000$ repeated simulations. These simulations ran in less than 60 minutes on a laptop with a 12th-generation Intel Core i7-1270P chip.

\begin{figure}[t]
\centering
\vspace{12.5pt}
\includegraphics[width=1.\columnwidth]{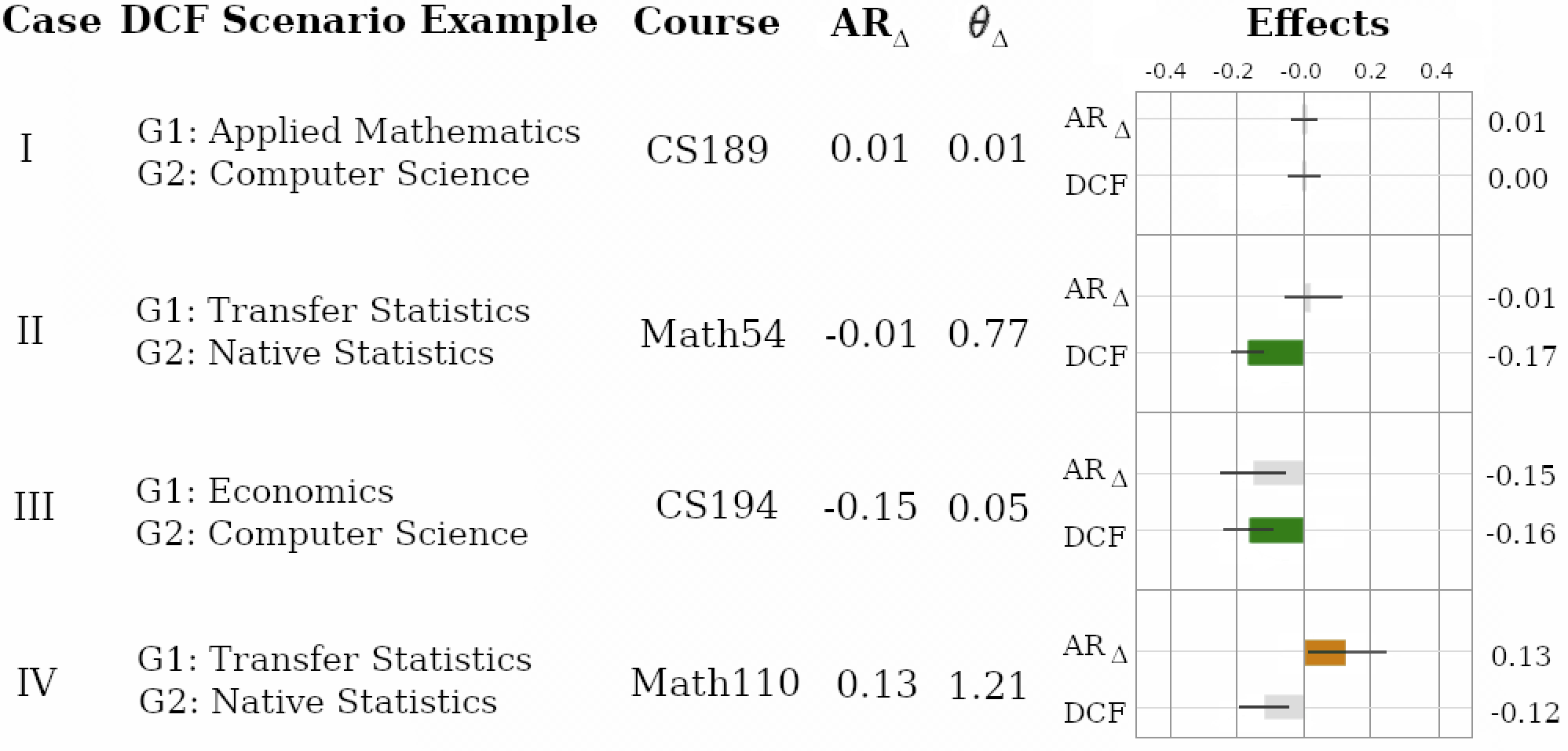}
\caption{Examples of courses illustrating the relationships between AR and DCF outlined in Figure~\ref{fig:dcf_pr_cases}. By decoupling course difficulty and student performance levels, the DCF methodology can yield more robust insights into disparities in course difficulties compared to analysis of AR differences.}
\label{fig:dcf_vs_pr_res}
\Description[]{long description}
\end{figure}

\section{Data and Preprocessing}
\label{sec:data}

\begin{table*}
    \centering
    \caption{Model selection results. For each studied major (and major pairing) dataset, we first identified the best-fitting IRT model based on the BIC criterion. Afterwards, we verified that the assumptions of the identified IRT model are fulfilled and that the model parameter fit is reliable and valid. The symbol ``+'' indicates whether the data also passes the assumption, reliability, and validity tests of the 1PL (Rasch) model underlying the DCF analysis.}
    \begin{tabular}{c|cc|cccccc}
       \multirow{2}{*}{Dataset} & \multirow{2}{*}{No. Students} & \multirow{2}{*}{No. Courses} & \multirow{2}{*}{Best BIC Score} & \multirow{2}{*}{Q3 Criterion} & \multicolumn{2}{c}{Reliability ($p<0.05$)}& \multicolumn{2}{c}{Validity ($p<0.05$)}\\
       \cline{6-9}
       &&&&&Random& Time & Student& Course \\
      \hline
        AppliedMath &                           1870 & 46   & 1PL       & \checkmark & $0.82$ & $0.86$ & $0.91$ & $0.57$\\
        Statistics&                             1047 & 53   & 1PL       & \checkmark & $0.70$ & $0.72$ & $0.91$ & $0.89$\\
        CompSci&                                5613 & 61   & 2PL 1DIM$^+$  & \checkmark & $0.63$ & $0.75$  & $0.92$ &$0.83$\\
        DataSci&                                2339 & 65   & 1PL       & \checkmark & $0.70$& $0.76$ & $0.96$ &$0.87$\\
        BusinessAdmin&                          4464 & 75   & 2PL 2DIM  & \checkmark & $0.64$ & $0.66$ &  $0.84$  & $0.83$\\
        Eco&                                    4742 & 64   & 1PL       & \checkmark & $0.66$& $0.68$ &$0.91$ & $0.96$ \\
        \hline
        BusinessAdmin $+$ Eco&             9166 & 119  & 2PL 2DIM  & \checkmark & $0.63$& $0.58$ & $0.86$ &$0.86$\\
        Eco $+$  CompSci &                 10234& 99   & 2PL 1DIM$^+$  & \checkmark & $0.74$ & $0.75$ & $0.90$ & $0.81$ \\
        Eco $+$  DataSci &                 6893 & 94   & 1PL       & \checkmark & $0.70$& $0.73$ &$0.88$ & $0.96$\\
        AppliedMath $+$  CompSci&          7191 & 68   & 2PL 1DIM$^+$  & \checkmark & $0.78$& $0.80$ &$0.84$ &$0.88$\\
        Eco $+$  Statistics&               5792 & 91   & 1PL       & \checkmark & $0.69$& $0.72$ & $0.89$ &$0.88$\\
    \end{tabular}
    \label{tab:selection_results}
\end{table*}

\subsection{Enrollment Data}

The present study leverages a large course enrollment and grade dataset collected between the Fall semester of 2011 and the Spring semester of 2022 at a large public university in the United States, capturing data from over $196{,}000$ students and $8{,}600$ unique courses. The unit of analysis in this dataset is a unique completed student course enrollment with a received letter grade on an A-F scale. These letter grades can be determined based on formative assessment throughout the semester, or on end-of-semester exams, or a combination of both, depending on the course. These letter grades can be represented numerically for IRT modeling, as described in Section \ref{sec:method:preproc}. We note that we excluded student enrollments who took courses for pass/fail ratings (about one third of the student course pairs), as establishing a joint scale between letter and non-letter grades is non-trivial, letter grades were the most common form of grading. The pass/fail courses at the study site were further known to disproportionately include courses taken for interest or leisure rather than degree attainment.

We investigate differential effects of course difficulty across major pairs and transfer status. For generating relevant major pairs, ensuring sufficient overlap in student course enrollment is crucial for a reliable parameter identification~\cite{haas2023bayesian}. Therefore, we determined the most common significant pairs based on the count of majors pairings in which students were enrolled. Further, we aimed to have at least one non-STEM degree among the pairs, which made us opt for the 7th most common major combination instead of the 5th most common. The resulting major pairs were (Business Administration $+$ Economics), (Economics $+$ Computer Science), (Economics $+$ Data Science), (Applied Mathematics $+$ Computer Science), and (Economics $+$ Statistics). After subsetting the grade dataset to these $6$ relevant majors, the numbers of students and courses were $23{,}366$ and $386$ respectively.

\subsection{Preprocessing}
\label{sec:method:preproc}
IRT expects dichotomous response types (i.e., $\{0, 1\}$), but the grades in the raw dataset are on an A-F scale. Thus, to fit the logistic IRT models, all grades must be transformed into dichotomous grades. Relatedly, a prerequisite for IRT-based analysis is variance in the course grades as a measure of information~\cite{haas2023bayesian}. If we chose course pass/fail as basis for the transformation, the dichotomous grade distribution would be imbalanced, leading to many courses showing no variance in the dichotomous outcomes due to all students achieving the passing grade. This is an obstacle to IRT parameter identification via maximum likelihood fitting. Following prior work~\cite{jiang2019goal}, we aimed to maximize variance in the dichotomous grade distribution by using the median of all grades as the achievement grade (AG). The median for our dataset is the letter grade A. Hence, we encode A grades with $1$ and all lower grades with $0$.

Further, we filter courses with few participants and students with few enrollments to ensure robustness of parameter estimates. We follow prior work~\cite{baucks24lak} and iteratively filter students with less than $5$ grades and courses with less than $20$ students until the dataset size remains constant. The resulting dataset employed in our analysis consists of $6$ majors with $20{,}075$ students and $364$ unique courses.

\subsection{Groupings}

\subsubsection{Major} At the study site, courses for students in different majors can overlap substantially (e.g., economics and business administration). Our analysis focuses on gaining insights into how students from different departments experience each other's courses. To investigate this, we group the students by major and evaluate potential DCF effects for each course.
    
\subsubsection{Transfer} Students transferring to the university is a common occurrence (approximately 6,000 in Fall of 2023). About 92\% of admitted transfer students arrive from in state community colleges.
To be considered for transfer, students must complete transfer courses, some of which are recognized for course credit at the study site (i.e., course articulation).
We are interested in whether post-transfer courses are equally challenging for transfer and native students. We assess this via DCF evaluation. Transfer students account for approximately 10\% of the students in the preprocessed dataset. 

Lastly, we open-source our analysis code via GitHub to facilitate the adoption of DCF as a CA methodology.\footnote{\href{https://github.com/FrederikBaucks/irt-course-evaluation}{https://github.com/FrederikBaucks/irt-course-evaluation}}

\section{Results}
\label{sec:results}
    \subsection{Verifying IRT Model Assumptions}

\begin{figure*}
    \centering
    \includegraphics[width=1\textwidth]{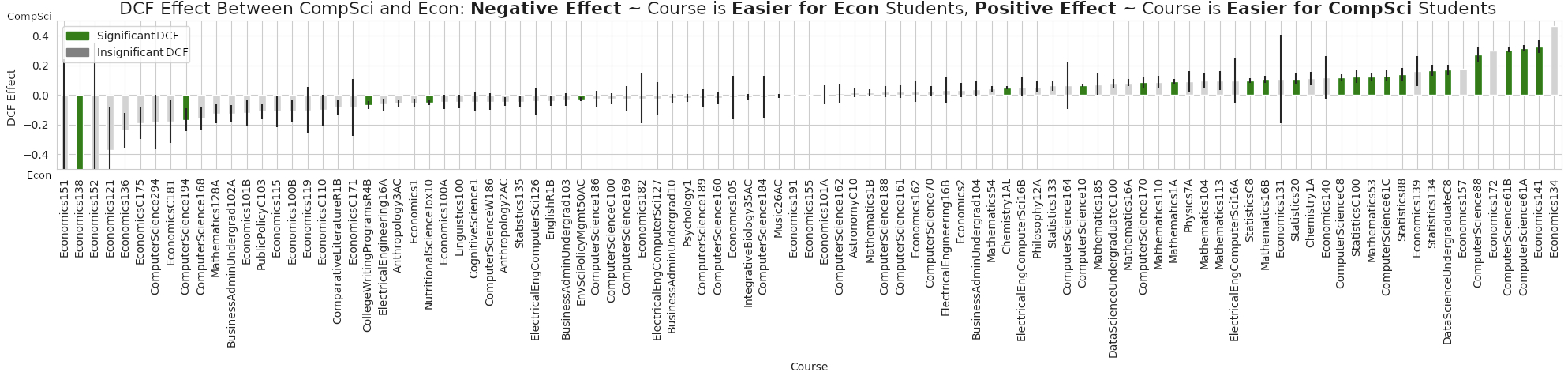}
    \includegraphics[width=1\textwidth]{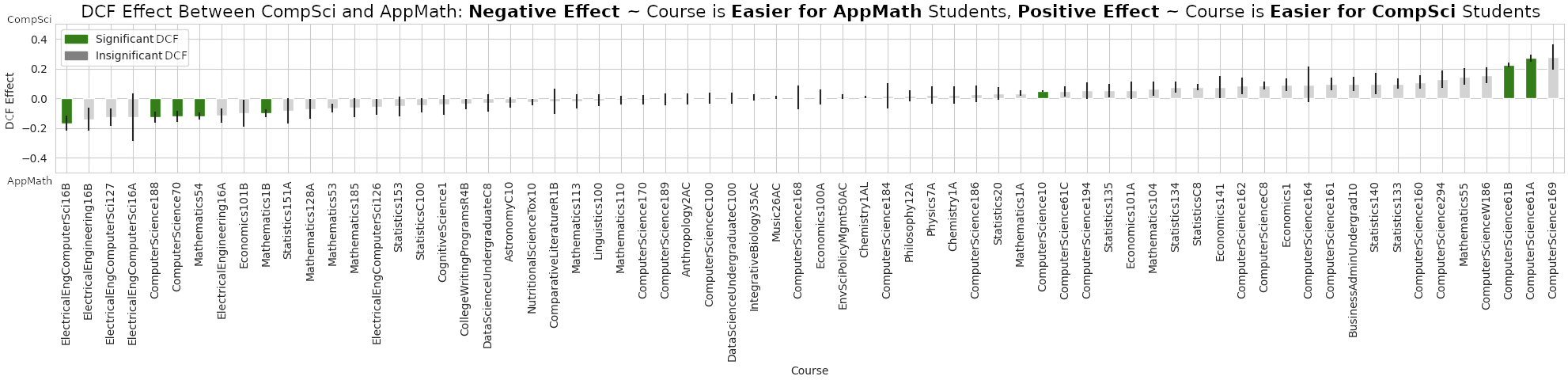}
    \caption{ 
    DCF effects for Major/Minor Pairings. [Top] Economics and Computer Science, [Bottom] Applied Math and Computer Science.  Course names include the department and a number abbreviation (e.g., ``Statistics140''). Lower division courses have a number less than or equal to 100. Otherwise, they are upper-division courses. Green bars indicate a significant effect. A negative effect means that the course is easier for the first major in the tuple, and positive effects are the opposite. The effect size is the percentage difference between the groups at the average student trait value across both groups. If there is an effect of $0.2$, the student in the preferred group with an average trait value of both groups is $20\%$ more likely to receive a response of $1$. 
    }
    \label{fig:dif_major_minor}
    \Description[]{long description}
\end{figure*}

Table~\ref{tab:selection_results} shows the results of model selection and testing of assumptions for the datasets related to the six individual majors, verifying that IRT models are well suited for the different academic fields, including STEM and non-STEM majors, and confirming the models' assumptions for large-scale data.

Firstly, we observe that, according to the BIC, a one-dimensional model yields the best fit for the majority of datasets (9/11). Exceptions are the major Business Administration and the major pair (Business Administration $+$ Economics), which correspond to the multidimensional 2PL-2DIM model.
The Q3 criterion, which can indicate if the local independence assumption is violated, shows that in none of the analyses are the scores in the data at risk of violating the assumption. This means that the main assumptions of IRT (dimensionality and local independence) are generally met (see Section~\ref{model_assumptions}). In addition, the reliability and validity assessments, represented by correlation coefficients, give us confidence that the fitted model parameters are robust and do represent student performance and course difficulty well. Thus, following the methodology, we conclude that IRT suits the different majors in our CA datasets. 

Since the DCF methodology (Section \ref{sec:methodology}) is based on the 1PL model, it is important to verify that the model assumptions for the 1PL model are also by by the data sets. We indicate this in Table \ref{tab:selection_results} using the ``+'' symbol in the BIC column. Only for the two datasets that include Business Administration, which the 2-dimensional 2PL model fits best, is IRT's assumption of unidimensionality by the remaining dimension in the 1PL model at risk. Because of this we exclude this major and the corresponding major pairing (Business Administration $+$ Economics) from the following DCF analyses.

\subsection{Evaluation of DCF Effects}
    
    \subsubsection{Comparing DCF and AR$_\Delta$}

        Following the methodology outlined in Section \ref{dcf_vs_pr}, we provide real world examples that illustrate the four hypothetical cases (I-IV) of DCF findings in Figure \ref{fig:dcf_vs_pr_res}. These examples highlight the relevance of DCF over AR-based measures when capturing differences between groups reliably. 
        Especially in cases II: $\theta_\Delta$ and IV: $\theta_\Delta +$DCF, where the mean student trait $\phi_{G_\cdot}$ differs between groups $G_1$ and $G_2$, the variations between the two effects can be particularly large. This is because DCF can account for the student trait difference, while the AR measure cannot. 

        \begin{figure*}[t]
            \centering
            \includegraphics[width = \textwidth]{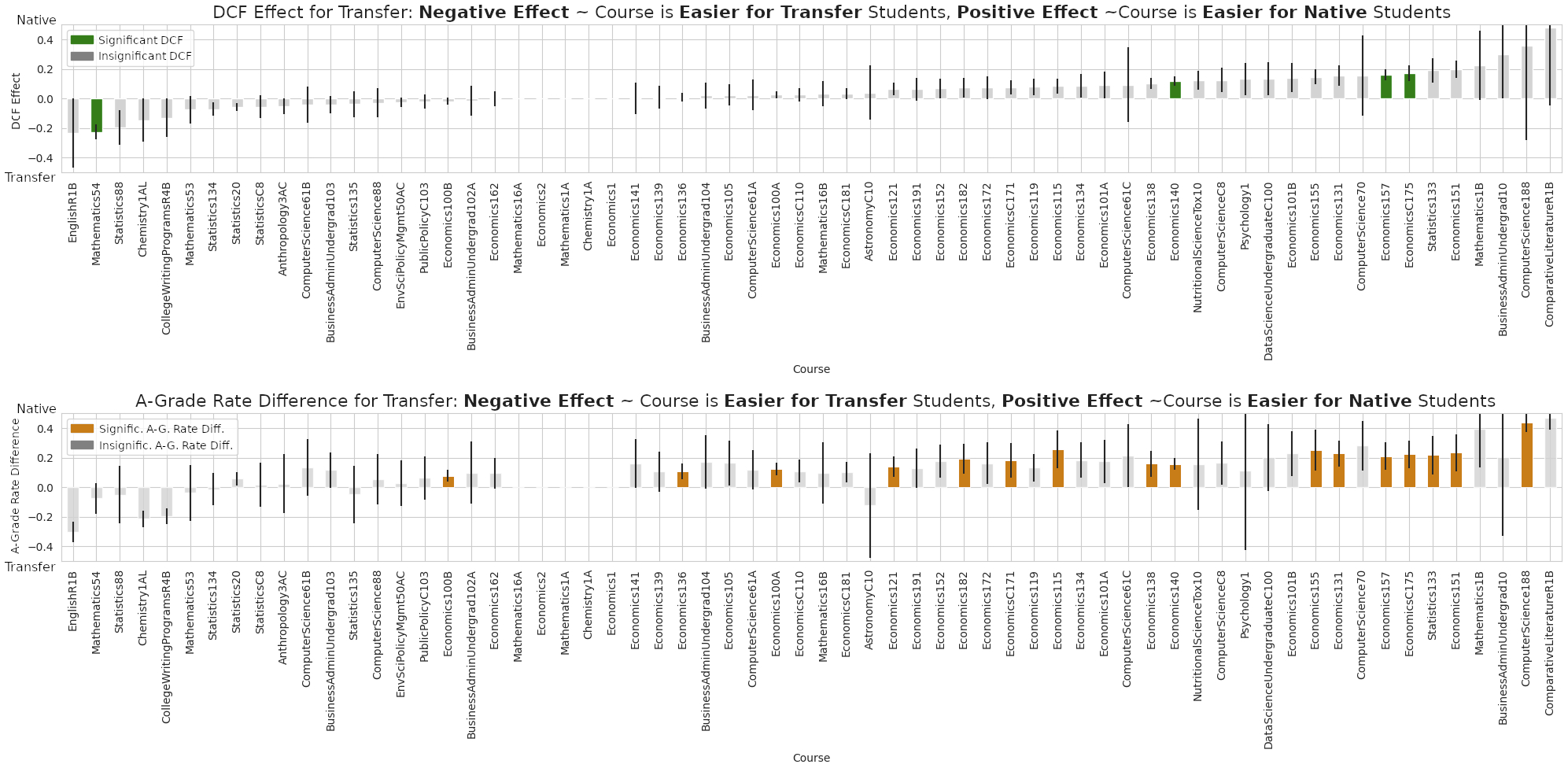}
            \caption{Comparison of [Top] DCF effects, similar to Figure \ref{fig:dif_major_minor}, and [Bottom] course AR group differences for Transfer Status of the Economics major. Again,  the DCF effect sizes ([Top]) are the differences in passing probability between the groups at the average student trait value across both groups. The results indicate that DCF is more robust than ARs due to mitigation of student performance. Differences in effect significance and effect size between DCF and ARs in the same courses indicate that DCF classifies far fewer effects as significant, and effect sizes can differ greatly between measures.}
            \label{fig:dif_transfer}
            \Description[]{long description}
        \end{figure*}
        
    \subsubsection{Major Affiliation DCF}
    Figure \ref{fig:dif_major_minor} visualizes the results of the DCF analysis for the majors Economics (Eco), Computer Science (CS), and Applied Mathematics (Math) using the two major pairings (Eco $+$ CS) and (Math $+$ CS).

    For the first major pairing (Eco $+$ CS) in Figure \ref{fig:dif_major_minor} [Top], significant DCF effects are highlighted in green and indicate, if positive, that course are easier for CS majors. Negative effects indicate that the courses are easier for Eco majors. More courses are significantly easier for CS students than for Eco students (18/23). In particular, the CS, Mathematics, and Statistics departments are frequently (16/17) among the courses favorable to CS students. The content overlap of courses in these departments suggests that there are course-specific factors between the different majors that we can measure with the DCF effect: CS students have an easier time in courses that are closer in content to their major, even after we account for student trait values. This observation is reinforced by the second major pairing (Math $+$ CS) in Figure \ref{fig:dif_major_minor} [Bottom]. Here, we find fewer courses with significant DCF effects that are easier for CS students (3/8). At the same time, Math and CS are closer to each other in terms of content, which is reflected by a smaller number of significant effects (8 in (Math $+$ CS) vs. 23 in (Eco $+$ CS)). 

    We can further observe a difference in the strength of the DCF effects. The DCF effect in Figure \ref{fig:dif_major_minor} is given as the difference in the AR probability under the average student trait of students in both groups combined. Specifically, given a student with an average trait value, we can identify courses that may be less aligned with students prior knowledge as elective courses for non-majors than others, e.g., ``ComputerScience10'' is more suited to the knowledge of Eco students than ``Statistics20'', when compared to CS students.
        
    \subsubsection{Transfer Status DCF}

    When grouping students by transfer status, we observe fewer significant DCF effects (at most 4 in the Eco major) compared to the previous major pairing analysis. This indicates that transfer and native students experience course difficulty similarly independent of their student trait value. In Figure \ref{fig:dif_transfer}, we show a comparison of DCF effects [Top] with AR$_\Delta$ effects [Bottom] for the Eco major. Significant effects are highlighted in color. Comparing DCF effects to AR$_\Delta$, the ARs of the two groups (Figure \ref{fig:dif_transfer} [Bottom]) show multiple significant differences in AR (17/48) which are not present in the DCF plot [Top] (4/48). This suggest that the differences in A-grade achievement between the two groups are not as often due to course-specific difficulty factors but rather due to variations in the average student trait values of transfer and native students, perhaps attributable to the greater variety of preparedness among transfer students.

    \subsection{Power Analysis}

\begin{figure}
            \centering
            \includegraphics[width = 0.95\columnwidth]{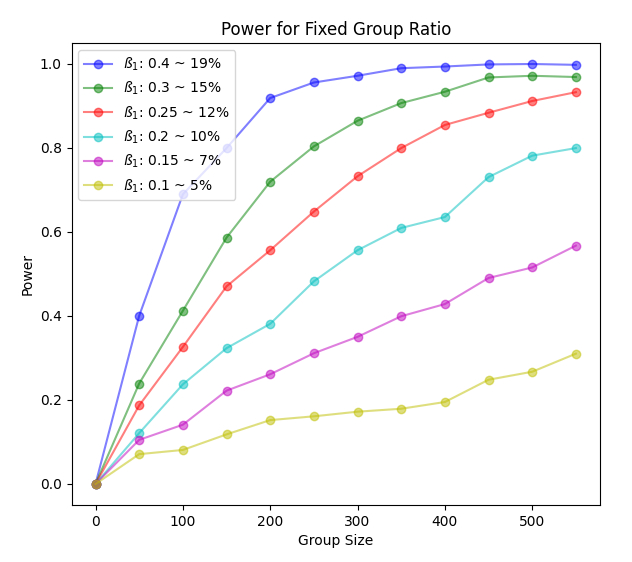}  
            \caption{Simulation study: $50$ courses, $1$ DCF course, and $1000$ simulation runs as fixed parameters. We show the power analysis of DCF under different DCF effects $\beta_1$ given as logits for all $\theta_{s\in S}$ and percentages at $\theta=0$ using $p = 0.05$. 
            We fix the group ratio to $0.5$ and vary group sizes simultaneously (x-axis) to reflect different dataset sizes. The results show sufficient power ($0.8$) for effect sizes larger than 0.2 ($\sim10\%$ achievement probability difference) and dataset sizes $\geq2 * 500$. Smaller datasets show sufficient power only for larger effects.
            }
            \label{fig:sim_power}
            \Description[]{long description}
            
        \end{figure}

        Following Section \ref{simulation_and_bh}, we conduct a simulation study and report power analyses results for our DCF detection methodology in Figure \ref{fig:sim_power}. The simulations estimate the statistical power (true discovery rate) of the DCF detection for different effect sizes and student population sizes.
        Comparing the DCF effect sizes in the simulation to the effect sizes detected in the real data (Figures \ref{fig:dif_major_minor} and \ref{fig:dif_transfer}), the DCF effects in the real data often exceed a $12\%$ point AR difference, which corresponds to $0.25$ on the logit scale. 
        Thus, we focus on the red and green curves in Figure \ref{fig:sim_power}, which represent $\beta_1$ values of $0.25$ and $0.3$, respectively. These curves indicate adequate power ($>0.8$) for group sizes greater than $>350$ for $\beta_1$ values of $0.25$ and $>250$ for $\beta_1$ values of $0.3$ under a fixed group ratio of $0.5$. Comparing these group sizes to our transfer status DCF analysis, we have an average course size of $583$ students. Thus, for a given group ratio of $0.5$, we obtain adequate power ($>0.8$) for a DCF effect size of $0.3$ in terms of logits or of $15\%$ points (green curve). 
        Since the major affiliation DCF examines pairs of majors, there are, on average, more students per course, i.e., $956$. Therefore, a DCF effect of $0.25$ logits, corresponding to $12\%$ points (red curve), is sufficient. 

        Together with the log-likelihood ratio test, we control our DCF detection method for false discovery rate (FDR) and true discovery rate (TPR).  
        This gives us confidence that our method has sufficiently low FDR and can reliably detect true DCF effects (statistical power). 
        
\section{Discussion}
\label{sec:discussion}

Capturing disparities in course difficulty across student groups, such as by departmental affiliation and transfer status, is an important challenge for curriculum analytics (CA) (e.g., ~\cite{fematt2021identifying,ishitani2008transfers}). Still, existing approaches for measuring such variations do not adequately decouple course-specific difficulty from students' general performance levels~\cite{baucks24lak}. The present study establishes differential course functioning (DCF) as an extension of item response theory (IRT) to overcome this limitation, enabling more nuanced and equitable insights into academic performance.

DCF detection identifies courses that vary significantly in difficulty for students with different group attributes (i.e., major affiliation). The number of courses that show significant differences deviates, sometimes significantly, from differences captured by achievement rate (AR) measures (see AR vs. DCF for transfer status in Figure~\ref{fig:dif_transfer}). 
In particular, we demonstrated through reliability and validity evaluations of the IRT parameters (Table~\ref{tab:selection_results}) that our DCF results are robust and meaningful.
This indicates that the differences between the AR measures and the DCF effects are due to the conceptual difference in measuring group differences.
As illustrated in the comparison between AR and DCF effects (Figure \ref{fig:dcf_vs_pr_res}), AR measures cannot account for student performance but are instead influenced by it (Figure \ref{fig:dif_transfer}). 
This limitation can lead to confounding and potential misinterpretation \cite{baucks21mitigating, ochoa2016simple}, and demonstrates the capability of DCF to uncover more nuanced effects. Existing CA methodologies for evaluating such group differences do not control for whether differences in course performance are due to \textit{local} course-specific factors (e.g., course content) or \textit{global} student-related factors (student traits).  
%
%
In particular, existing CA measures of course difficulty based on pass rates can lead to interpretation bias \cite{baucks21mitigating}, because they are confounded by course difficulty and students participating\cite{baucks24lak}.
Attempts to extract student achievement from AR measures (e.g., \cite{ochoa2016simple, molontay2020characterizing}) are also limited to separate calculations for group achievement levels (e.g., dividing students into three performance groups). In contrast, our DCF method provides a continuous approach that eliminates the need for group labeling and is more accurate.

The present study's findings suggest that IRT models used for DCF detection are suitable for fitting the grade data from higher education institutions in the US, extending prior work on German university samples \cite{baucks24lak}. We find evidence from model selection (BIC) and model verification (Q3, reliability, and validity) results that IRT's key assumptions (i.e., dimensionality and local independence) are met and that it is an appropriate model family for analyzing grade data across academic majors.
Except for ``BusinessAdmin'', the data failed to meet the unidimensionality assumption as a multidimensional trait model was required. 
Extending the DCF method from uni- to multidimensional traits is an exciting avenue for future work.
%
However, our analysis is restricted to a single institution in the US, so further replication to more institutions is desirable.

DCF detection can assess course-specific difficulty factors related to departmental affiliations by analyzing students' grades from various majors.
We observe significant DCF effects favoring courses mostly taught in the home department or strongly related in content to the student's major. Academic preparation can impact performance \cite{soares2009academic}. For instance, computer science (CS) majors perform better in statistics courses than economics majors, possibly due to their robust training in quantitative and computational thinking. This observation is echoed when comparing CS majors to applied mathematics majors. High-grade achievement in theory-intensive (e.g., differential equations) electrical engineering courses (i.e., ``ElectricalComputerSci16B'' and ``ElectricalEngineering16A'') are easier for applied mathematics majors. Thus, illustrating the impact of a student's major-specific academic preparation and foundational skills on their performance in courses across departments (e.g., ``Economics141'', as a statistics-heavy course, is easier for CS students). These insights can be utilized by advisors and policymakers to (i)~identify courses that may require additional preparation by students of a particular group and thus suggest specific summer courses to catch up or adjust the course sequence to include more preparatory instruction, and (ii)~use the magnitude of the effects to identify upper-division courses that may be more suitable for non-home department students than others (e.g., due to lower specialization). This approach improves curriculum relevance and aligns with efforts to personalize educational pathways (e.g., \cite{Auvinen2014stops, wagner2022combined}), ensuring they are equitable and responsive to diverse student needs. DCF effects related to major affiliation offer valuable insights for CA stakeholders such as student advisors and curriculum policymakers.

Transfer status-related DCF detection helps to understand the impact of the transfer process on students, controlling their grade achievement level. Detection reveals few significant DCF effects in our data set, suggesting that in most courses, transfer and native students experience similar course difficulty. Courses that indicate effects tend to be easier for transfers if the course is in the lower-division (course No. $\leq 100$) and easier for natives if it is in the upper-division (course No. $>100$). Lower-division courses (e.g., ``Mathematics54'', a linear algebra course) that are easier for transfer students may have content overlaps with courses frequented at students' prior academic institutions. When transfer students enter the university, the nominal academic year after transfer is $3$. Thus, students usually enroll in upper-division courses after transfer. Lower-division courses may substantially overlap with courses taken at the previous college, resulting in DCF effects since not all of a student's courses are necessarily transferable. Upper-division courses that are more difficult for transfer students may show lower levels of preparation or a misalignment in the prior curriculum. Articulation officers deciding on course transferability can use our DCF methodology to narrow transfer groups to single community colleges to measure college-specific challenges and support data-driven articulation decisions. Yet, these effects are related to the ability to achieve an A-grade and are unrelated to students' success atgraduating. On the study side, graduation rates of transfer and native students are similar, with transfers attaining a 91\% graduation rate and 92\% for natives. These insights advance our understanding of transfer students' challenges and highlight the critical need for tailored support strategies that bridge preparatory gaps and foster academic success across students of different backgrounds.

IRT allows separating student performance and course difficulty factors~\cite{Ayala2013:Theory}. Further, DCF enables the critical inspection of significant disparities in course-specific difficulty between different student groups, especially after controlling for their performance level as measured by the IRT traits. 
Using DCF detection, we uncover disparities that inform stakeholders, including advisors, articulation officers, and academic policymakers. 
%
Stakeholders can use DCF effects to distinguish between course-specific and group-specific effects. The results can support interventions that target disparate difficulties throughout a student's career. 
These difficulties may be due to a student's lack of preparation or knowledge prior to transfer. This benefits equity and fairness by, e.g., increasing course preparation and provides a rationale for follow-up analyses addressing unintended course disparities and investigating their causes.

\section{Limitations and Future work}
\label{sec:future_work}

IRT-based DCF detection relies on existing performance data and retrospectively assesses disparities in course-specific difficulty between distinct student groups. Thus, the method cannot make inferences about students and courses for which no data is available. Future work will investigate course representation learning methods such as Course2Vec \cite{pardos2019data} to predict the difficulty of new courses based on alternative data types (e.g., course descriptions). Predictive models learned from enrollment data could generalize course difficulty estimates at scale and elevate further insights into systematic differences in course difficulty by student group. 

The IRT models utilized in this study assume dichotomous grades. As a result, we converted letter grades (A-F) according to the median grade (A) to maintain variance in the modeled outcome. This implies that the student trait in the IRT-related DCF detection captures the ability of students to achieve a grade of A on their first attempt. This trait may be more stable than a student trait focused on another criterion, such as pass/fail, and describes a subset of student performance. Similarly, statements about course difficulty are limited to this discrimination of grades.
Thus, the traits should be interpreted cautiously, especially when making statements about the fairness of courses using DCF detection. For transfer and native students, we found that most A-rate differences are due to overall differences in the two groups' trait levels. However, we want to emphasize that in terms of graduation rates, transfer and native students are comparably successful, with transfer students achieving a four-year graduation rate of 91\%, closely mirroring the six-year graduation rate of 92\% for native students. In the future, we intend to use rating scale models~\cite{Ayala2013:Theory} that can leverage student achievement data of finer granularity (beyond dichotomous grades--A or below A) employing the full letter grade scale, that is, A-F. This enables better targeting of students with lower grades and students in grade-level groups, e.g., dropouts vs. non-dropouts.

We assume that courses are time-invariant in difficulty. Therefore DCF effects remain constant over time in a course, which may not always be true when applying IRT and DCF detection to college course data collected over multiple years~\cite{baucks24lak}. We have accounted for this for individual students using the time-dependent split-half reliability test, which indicates robust results. For courses, we chose time-invariance to capture global trends in DCF rather than semester-dependent trends. First, we believe stakeholders are interested in global interventions that target courses with a greater diversity of students. Second, based on the simulation study results, we believe that a course-based view of DCF effects would likely run into problems of statistical significance, requiring certain sample sizes (see Figure \ref{fig:sim_power}). Some courses, such as major/minor combinations, have limited overlap, making it difficult to draw conclusions. Applying DCF detection to course offerings with sufficient enrollment data will be an exciting avenue for future work and could, for example, indicate differences in teaching from semester to semester.

Lastly, eliciting students' perceptions of course difficulty has effectively predicted a course's challenge level before completion, as these correlate with final grades \cite{england2019student}. This approach could shed light on the workload associated with a course and help us understand the reasons for differences in course difficulty between groups of students and, if unintended, mitigate differences through active teaching efforts \cite{andres2019active}. We believe that the cognitive effort required by a course is crucial. Thus, analysis of how time commitment and stress contribute to difficulty may provide additional insights \cite{borchers2023insights}. This will lead to a better understanding of what makes a course difficult for student groups and how to effectively address this.

\section{Summary And Conclusion}
\label{sec:conclusion}
The present study introduced differential course function (DCF) as an extension of prior item response theory (IRT)-based methodologies in curriculum analytics (CA). DCF detection tests significance and quantifies the magnitude of disparate difficulty in courses experienced by students of different backgrounds while controlling for student performance. Thus, DCF provides insights into how different student groups experience challenges posed by individual courses to inform interventions promoting equity and fairness. We validated our methodology on institutional records of course grade data from a large public university in the United States. We found significant course difficulty effects (i) for majors, in which minor courses are challenging, and (ii) for transfer students, there were very few DCF effects, but when significant DCF was detected, transfers experienced lower division courses easier and upper division courses harder. Overall, DCF poses a promising direction for higher education stakeholders in monitoring, detecting, and intervening in course difficulty differences across academic subpopulations.

\subsubsection*{Acknowledgements} 

The work of FB and LW was supported by the Ministry of Science (NRW, GER) as part of the project KI.edu:NRW.

\bibliographystyle{ACM-Reference-Format}
\bibliography{bibliography}

\end{document}